\documentclass[a4paper,fleqn]{cas-sc}

\usepackage[numbers]{natbib}
\usepackage{hyperref}

\def\tsc#1{\csdef{#1}{\textsc{\lowercase{#1}}\xspace}}
\tsc{WGM}
\tsc{QE}

\begin{document}
\let\WriteBookmarks\relax
\def\floatpagepagefraction{1}
\def\textpagefraction{.001}

\shorttitle{GCtx-UNet}    

\shortauthors{Alrfou and Zhao}  

\title [mode = title]{GCtx-UNet: Efficient Network for Medical Image Segmentation}  



%

\author[1]{Khaled Alrfou}
\cormark[1]


\ead{kalrfou@uwm.edu}


\credit{Developed the GCtx-UNet method, conceived and designed the study, developed the software, evaluated
results, provided datasets, and contributed to the formal analysis and writing of the original draft.}

\affiliation[1]{organization={Electrical Engineering and Computer Science,University of Wisconsin-Milwaukee},
            addressline={3200 North Cramer Street}, 
            city={Milwaukee},
            postcode={53211}, 
            state={WI},
            country={USA}}

\author[1]{Tian Zhao}


\ead{tzhao@uwm.edu}


\credit{evaluated the results, contributed to the formal analysis and writing of the original draft, and proofread and reviewed the final manuscript}


\cortext[1]{Corresponding author}



\begin{abstract}
Medical image segmentation is crucial for disease diagnosis and monitoring. Though effective, the current segmentation networks such as UNet struggle with capturing long-range features. More accurate models such as TransUNet, Swin-UNet, and CS-UNet have higher computation complexity. To address this problem, we propose GCtx-UNet, a lightweight segmentation architecture that can capture global and local image features with accuracy better or comparable to the state-of-the-art approaches. GCtx-UNet uses vision transformer that leverages global context self-attention modules joined with local self-attention to model long and short range spatial dependencies. 
GCtx-UNet is evaluated on the Synapse multi-organ abdominal CT dataset, the ACDC cardiac MRI dataset, and several polyp segmentation datasets. In terms of Dice Similarity Coefficient (DSC) and Hausdorff Distance (HD) metrics, GCtx-UNet outperformed CNN-based and Transformer-based approaches, with notable gains in the segmentation of complex and small anatomical structures. Moreover, GCtx-UNet is much more efficient than the state-of-the-art approaches with smaller model size, lower computation workload, and faster training and inference speed, making it a practical choice for clinical applications.
\end{abstract}



\begin{keywords}
 \sep GC-ViT \sep U-Shaped \sep Medical image segmentation
\end{keywords}

\maketitle

\section{Introduction}\label{in}
Automated medical image segmentation is critical in providing valuable information for the prevention, diagnosis, progression monitoring, and prognosis of various diseases, as well as quantitative pathology assessment. The U-shaped deep-neural networks, which include encoders, decoders, and skip connections, are now the most widely used methods for medical image segmentation. Although the U-shaped networks have achieved state-of-the-art performance in numerous medical image segmentation tasks, it still has limitations. One primary limitation is the encoders' ability to effectively extract and integrate long-range and local features. Methods based on Convolutional Neural Networks (CNNs) such as UNet~\cite{ronneberger2015u} and UNet++~\cite{zhou2019unet++} excel at capturing local features, but they struggle to model long-range dependencies within data. While Transformer-based methods such as Swin-UNet~\cite{cao2022swin} can model long-range pixel relations, they lack spatial induction bias in modeling local information, which leads to unsatisfactory results. 

Past research explored CNN-Transformer hybrid architectures such as TransUnet~\cite{chen2021transunet} to capture global and local information but these models often significantly increase the number of parameters. This, in turn, translates to higher computational complexity, potentially limiting their practical applications. Recently, Hatamizadeh et al.~\cite{hatamizadeh2023global} proposed a Global Context Vision Transformer (GC-ViT) that leverages global context self-attention modules and is joined with local self-attention to effectively and efficiently model both long and short-range spatial interactions. 
GC-ViT achieved state-of-the-art results across image classification, object detection, and semantic segmentation tasks.

In this paper, we introduce GCtx-UNet, a UNet-like segmentation network designed for medical image segmentation. GCtx-UNet effectively captures both long and short-range semantic features using GC-ViT~\cite{hatamizadeh2023global} encoders and decoders with skip connections. This architecture enhances performance while requiring fewer model parameters, with higher inference speed, and lower computational complexity. 

We evaluated the segmentation and runtime performance of GCtx-UNet on several medical image datasets including Synapse, ACDC, and several polyp image datasets. Our experimental results show that GCtx-UNet has better or comparable performance than the state-of-the-art segmentation algorithms including CNN-based, Transformer-based, and hybrid segmentation networks. Also, GCtx-UNet has the smallest model size and uses the least amount of training time and inference time. In addition, we pre-trained GCtx-UNet on both ImageNet and MedNet, which is a set of 200,000 medical images collected from public sources. GCtx-UNet pre-trained on in-domain images (i.e. MedNet) yielded better accuracy than GCtx-UNet pre-trained on natural images (i.e. ImageNet).

The source code for the model is available at \href{https://github.com/Kalrfou/GCtx-Unet}{github.com/Kalrfou/GCtx-Unet}.

\section{Related work}

The CNN-based methods are widely used and regarded as one of the most prominent approaches for medical image segmentation. Convolutional neural networks (CNNs), particularly encoder-decoder based architectures like UNet~\cite{ronneberger2015u} and its derivatives, have shown exceptional efficacy in medical image segmentation. For instance, Att UNet~\cite{oktay2018attention} enhanced segmentation through attention gates while UNet++~\cite{zhou2019unet++} 
introduced an alternative skip connection mechanism, nested and dense, alleviating the semantic gap between levels of UNet to a certain degree. This modification yields notable performance improvements compared to UNet. However, UNet++ cannot capture semantic features at full scale. Huang et al.~\cite{huang2020unet} proposed UNet3+ to maximize the use of full-scale feature maps by combining low-level details from various scales with high-level semantics. 
CNN-based methods have found application in diverse medical image segmentation tasks, such as retinal image segmentation~\cite{fu2022deau} and skin segmentation~\cite{zhang2023accpg}, showcasing promising performance and practicality in implementation and training.
Segmentation algorithms based on ResNet architecture have established its presence in medical image segmentation~\cite{li2024enhanced}. For example, Res-UNet~\cite{xiao2018weighted} enhanced retinal vessel segmentation with a weighted attention mechanism.

The self-attention mechanism (MSA) inherent in Transformers empowers them to perform global correlation modeling, enabling them to handle long-range dependencies effectively. Leveraging this capability, Transformers have made significant strides in both natural language processing and computer vision tasks due to their superior global modeling abilities.
Several pioneering studies have introduced Transformer-based architectures for medical image segmentation. Cao et al.~\cite{cao2022swin} presented Swin-UNet, integrating a Swin Transformer~\cite{liu2021swin} into a U-shaped segmentation network for multi-organ segmentation. Azad et al.~\cite{azad2022transdeeplab} proposed TransDeepLab for skin lesion segmentation, enhancing DeepLab with diverse window strategies. Additionally, Huang et al.,~\cite{huang2022missformer} introduced MISSFormer to leverage global information across different scales for cardiac segmentation, while Azad et al.~\cite{azad2023enhancing} introduced TransCeption, refining the patch merging module to capture multi-scale representations within a single stage. Combining convolution operations with a Transformer on the encoder side, Transclaw UNet~\cite{chang2021transclaw} enables detailed segmentation and long-distance relationship learning. UNETR~\cite{hatamizadeh2022unetr} adopts sequence-to-sequence prediction for 3D medical image segmentation.
These developments underscore the transformative impact of Transformer-based approaches on medical image segmentation, charting a path towards broader adoption and deep learning advancement.

Swin Transformer~\cite{liu2021swin} introduced local-window-self-attention to reduce the cost so that it grows linearly with the image size, used shifted-window-attention to capture cross-window information, and exploited multi-resolution information with hierarchical architecture. However, the shifted-window-attention struggles to capture long-range information due to small coverage area of shifted-window-attention and lacks inductive bias like ViT~\cite{dosovitskiy2020image}. 

\begin{figure}[ht!]
\centering
\includegraphics[scale=.50]{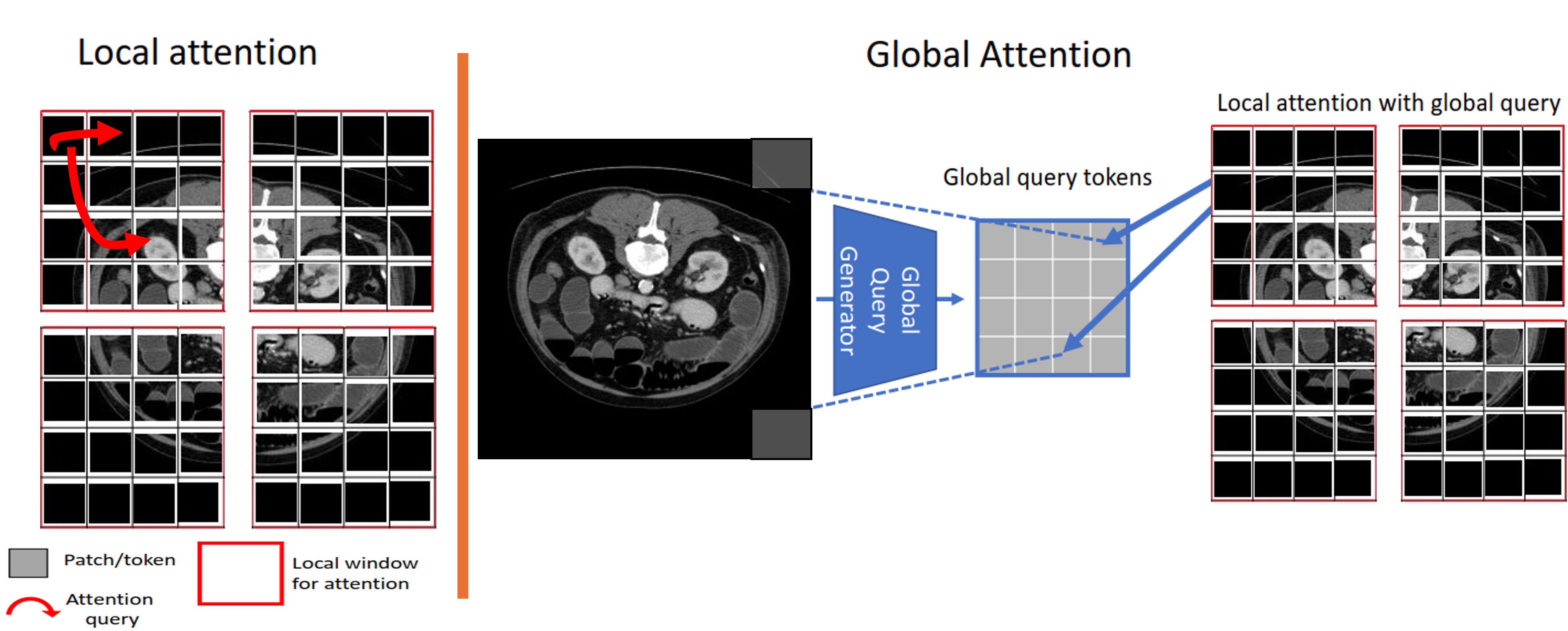}
\caption{An illustration of the local and global attention mechanisms in GC-ViT~\cite{hatamizadeh2023global}. Local attention is computed on feature patches within local window only (left). The global attention mechanism extracts query patches from the entire input feature map, aggregating information from all windows. The global query is interacted with local key and value tokens, hence allowing to capture long-range information.}\label{fig1}
\end{figure}

GC-ViT~\cite{hatamizadeh2023global} is a hierarchical architecture like Swin Transformer but utilizes global-window attention instead of shifted-window attention for effectively capturing long-range information. GC-ViT also uses convolution layers for downsampling to provide the network with desirable properties such as locality bias and cross-channel interactions which are missing in both ViT and Swin Transformer.
GC-ViT has 4 stages, each of which consists of alternating blocks of local and global Multi-head Self-Attention (MSA) layers.
As shown in Figure~\ref{fig1}, at each stage,  global query tokens are computed by using novel fused inverted residual blocks that encompass global contextual information from different image regions. While the local self-attention modules are responsible for modeling short-range information, the global query tokens are shared across all global self-attention modules to interact with local key and value representations. 

\section{Pre-training on MedNet dataset}
The majority of CNN and Transformer-based segmentation models are pre-trained on natural images such as ImageNet. However, this is suboptimal for medical image segmentation due to the semantic gap between natural and medical image modalities~\cite{shamshad2023transformers,alrfou2024cs}. In this work, we pre-trained the GC-ViT~\cite{hatamizadeh2023global} model, specifically GCVit xxTiny, on a large medical image dataset called {\em MedNet} that contains more than 200,000 medical images collected from several public datasets~\cite{subramoniam2022deep} and Kaggle~\cite{bib12,bib13,bib14}.

MedNet consists of different types of microscopy images such as X-ray, computed tomography (CT), optical coherence tomography (OCT), and MRI. Images in MedNet are divided into 65 classes. Similar to the approach of Stuckner {\em et al.}~\cite{stuckner2022microstructure} and Alrfou {\em et al.}~\cite{alrfou2024cs}, the MedNet dataset is divided into training and validation sets, with each class having 100 images in the validation set, resulting in 96.75\%/3.25\% training/validation split. Using 100 images per class for validation is sufficient to obtain reliable accuracy metrics and to prevent overfitting during training. Although the validation sets are balanced, the training sets exhibited some class imbalance. There are a few classes, each of which contains less than 0.12\% of the total images. Three classes contain 6.2\% of the images. 
Most classes have over 2000 images representing one to two percent of the training set. MedNet includes images from various modalities such as X-ray, CT, OCT, and MRI, and encompassed a wide range of medical diseases such as Kidney Cancer, Cervical Cancer, Alzheimer's, Covid-19, Pneumonia, Tuberculosis, Monkeypox, Breast Cancer, and Malaria.

We trained and tested GC-ViT xxTiny with the AdamW optimizer~\cite{kingma2014adam} for 100 epochs with an initial learning rate of 0.0001, weight decay of 0.05, and cosine decay scheduler.
The training data had been augmented using the albumentations library, which included random changes to the contrast and brightness, vertical and horizontal flips, photometric distortions, and added noise.

The training process continued until the validation score showed no improvement, employing an early stopping criterion with a patience of 10 epochs. Performance was evaluated using top-1 and top-5 accuracy metrics. Top-1 accuracy measures the percentage of test samples for which the correct label is the top prediction, while top-5 accuracy measures the percentage of test samples for which the correct label appears within the top five predictions.  
The top-1 accuracy of the GC-ViT xx-Tiny model is 82.3\%, and the top-5 accuracy is 98.2\%. 


\section{GCtx-UNet Architecture}
\label{meth}

\begin{figure}[ht!]
	\centering
	\includegraphics[scale=.4]{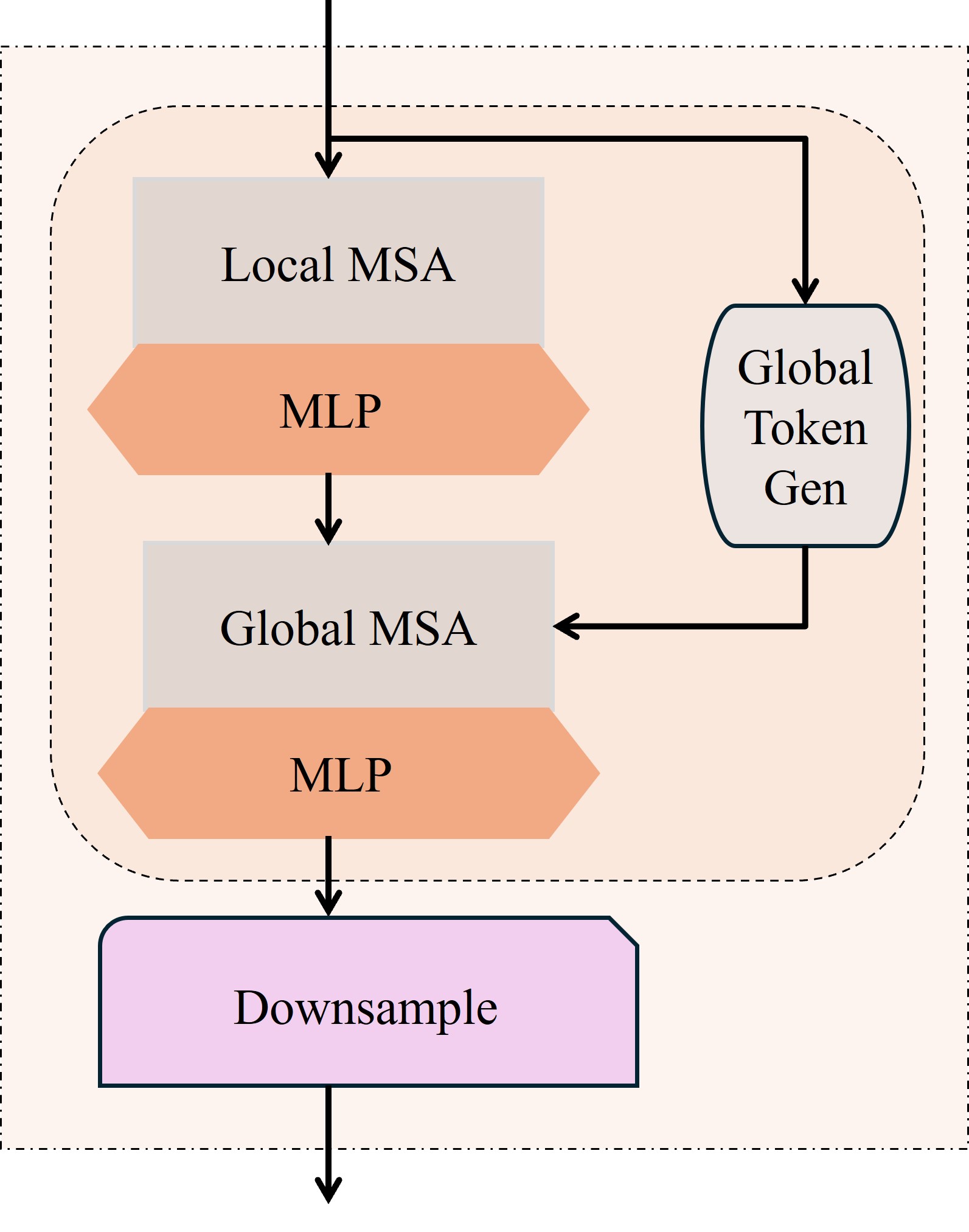}
	\caption{A GC-ViT block has a local and global attention, a global token generator, and a downsampling layer.}\label{fig_gcvitblock}
\end{figure}

The core component of GCtx-UNet is GC-ViT block, the local and global attention mechanisms of which are illustrated in Figure~\ref{fig1}. 
As shown in Figure~\ref{fig_gcvitblock}, Each GC-ViT block includes a local and global Multi-head Self-Attention (MSA), Multilayer Perceptron (MLP), a Global Token Generator (GTG) and a downsampling layer. The GTG component adds global context to the computations.
Local MSA can only query patches within a local window, while global MSA can query different image regions while still operating within the window. At each stage, the global query component is pre-computed.
The block also introduces a CNN-based module in the downsampling layer to include inductive bias, a useful feature for images that have been missing in both ViT and Swin Transformer.

\begin{figure}[]
	\centering
	\includegraphics[scale=.50]{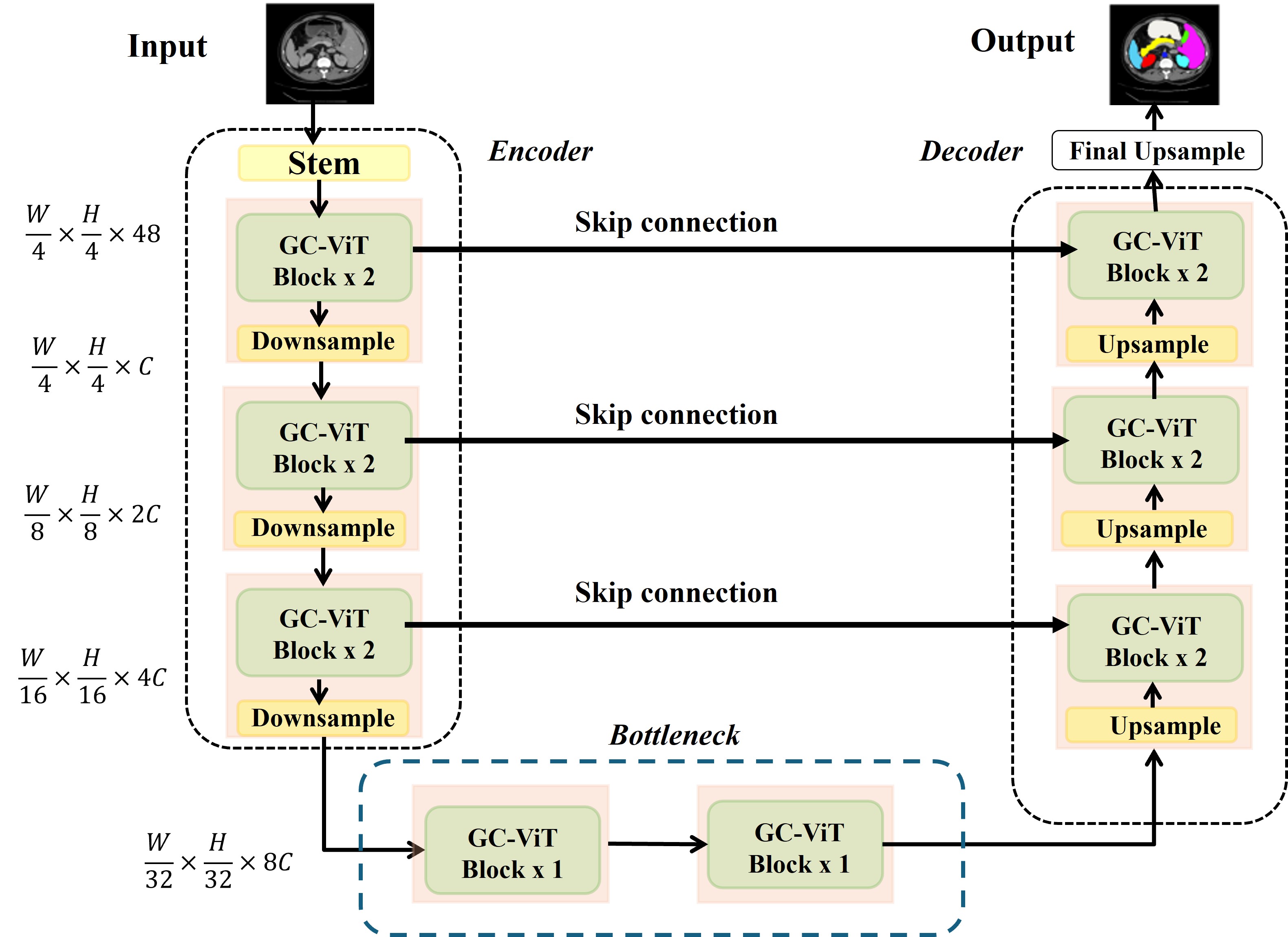}
	\caption{GCtx-UNet architecture includes encoders, bottlenecks, skip connections, and decoder. Encoder, bottleneck and decoder are all
constructed based on GC-ViT block}\label{GCtx-Unet}
\end{figure}

GCtx-UNet is a GC-ViT-based U-shaped Encoder-Decoder architecture with skip-connections for long and short-range semantic feature learning. As shown in Figure~\ref{GCtx-Unet}, the GCtx-UNet consists of encoder, bottleneck, decoder, and skip connections. 
 
\begin{enumerate}
    \item Both encoder and decoder used GC-ViT~\cite{hatamizadeh2023global} to model long and short-range spatial interactions, without the need for expensive operations such as computing attention masks or shifting local windows.
    \item At each stage, the GC-ViT encoder and decoder consist of alternating local and global self-attention modules to extract spatial features. Both operate in local windows like Swin Transformer.
    \item Skip connections concatenate the feature maps from the GC-ViT encoder with the corresponding decoder stages. The bottleneck is employed to acquire the deep feature representation, maintaining both feature dimension and resolution unchanged within this component.
    \item The downsampler between stages in the encoder part and upsampler between stages in the decoder part provide desirable properties such as inductive bias and modeling of inter-channel dependencies.
\end{enumerate}
 
Within the encoder, the initial image is partitioned into four patch blocks, which act as input for the four-stage GC-VIT module. Following the encoding process, the image dimensions are decreased to (H/32) × (W/32). In the decoder, the upsample operations are utilized to increase the image dimensions by 2 and reduce the number of channels by 2. The features from each stage of the encoder are concatenated with their corresponding stage in the decoder using skip connections. The decoder accomplishes its task through upsampling. 


\subsection{Encoder}
The encoder utilizes a hierarchical GC-ViT approach to acquire feature representations at various resolutions. This is achieved by reducing spatial dimensions while simultaneously increasing embedding dimensions by a factor of 2 across 4 stages. Initially, the input image $x \in \mathbb{R}^{H \times W \times 3}$ undergoes processing through the patchify layer. This layer comprises a $3 \times 3$ convolution operation with a stride of 2, along with padding, to generate overlapping patches. Subsequently, these patches are projected into an embedding space of dimension $C$ via another $3 \times 3$ convolution layer. After each stage in the GC-ViT backbone, the spatial resolution is decreased while the number of channels is increased through a downsampling layer. This downsampling operation helps in extracting hierarchical features at different resolutions.


\subsection{Downsampler}
In the downsampler, we incorporate the Fused-MBConv module to
generates hierarchical representations by injecting inductive bias into the network and modeling inter-channel correlations. Then, the convolution layer with a kernel size of 3 and a stride of 2 was used to downsample the spatial feature resolution by 2 while doubling the number of channels.
\begin{figure}[]
	\centering
	\includegraphics[scale=.4]{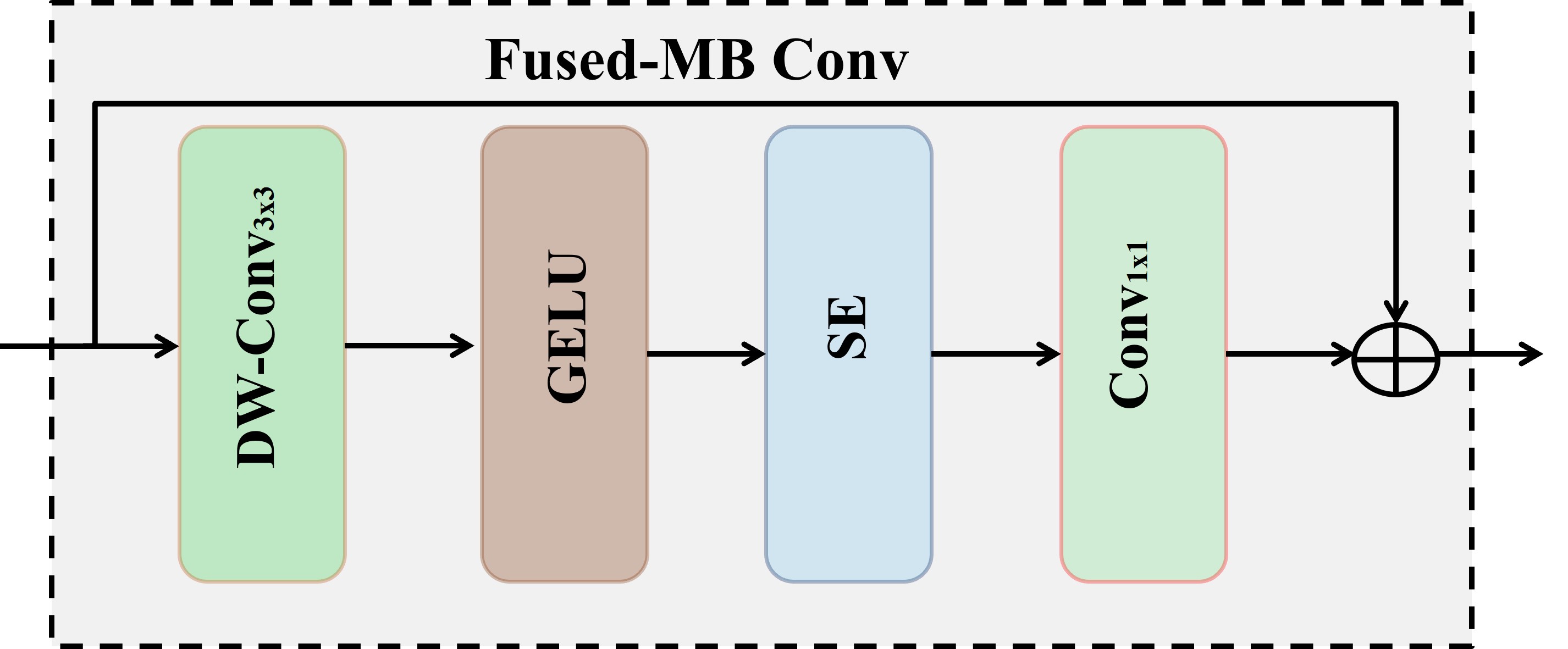}
	\caption{Fused-MBConv module}\label{usedMB}
\end{figure}
The Fused-MBConv, as shown in Figure~\ref{usedMB}, includes  DW-Conv$_ {3\times3}$, GELU, SE, and Conv$_ {1\times1}$.
The Fused-MBConv operation can be defined by the following equations:
\begin{gather*}
\hat{x} = \text{DW-Conv}_{3 \times 3}(x), \\
\hat{x} = \text{GELU}(\hat{x}), \\
\hat{x} = \text{SE}(\hat{x}), \\
x = \text{Conv}_{1 \times 1}(\hat{x}) + x.
\end{gather*}
where DW-Conv refers to Depthwise-CNN, SE refers to the Squeeze and Excitation block, and GELU represents the Gaussian Error Linear Unit function.

\subsection{Bottleneck}
Similar to Swin-UNet~\cite{cao2022swin}, two  GC-ViT blocks are used for bottleneck construction. The bottleneck is strategically designed to facilitate the learning of deep feature representations. Within this structure, the feature dimension and resolution remain unchanged.


\subsection{Decoder}
The symmetric decoder, corresponding to the encoder, is constructed using the GC-ViT Transformer block. The decoder mirrors the encoder design, replacing the patchy block with an unpatched block, the embedding layer with a de-embedding layer, and the downsample block with an upsample block.
%
The decoder's upsample block replaces the encoder's downsample block. The upsample block restructures the feature map of adjacent dimensions into a higher-resolution feature map and reduces the feature dimension by half. This upsample block effectively increases the spatial resolution while refining and normalizing feature representations, making it suitable for decoding and reconstructing higher-resolution features in segmentation models.
%
%
The skip connection fuses the features of the encoder with the deep features recovered from the up-sample, therefore mitigating the loss of spatial data produced by the downsampling.

\begin{table}[ht!]
\caption{List of the methods that used in our experiments.}
\begin{tabular}{|l|l|l}
\cline{1-2}
{\bf Architecture} & {\bf Method} &  \\ \cline{1-2} 
\multirow{6}{*}{CNN} & U-Net~\cite{ronneberger2015u} &  \\ \cline{2-2}
 & Att-UNet~\cite{oktay2018attention} &  \\ \cline{2-2}
 & R50-UNet~\cite{chen2021transunet} &  \\ \cline{2-2}
 & R50-AttUNet~\cite{chen2021transunet} &  \\ \cline{2-2}
 & UNet++~\cite{zhou2018unet++}&  \\ \cline{2-2}
 & PraNet~\cite{fan2020pranet} &  \\ \cline{1-2}

\multirow{2}{*}{Transformer} & Swin-UNet~\cite{cao2022swin}&  \\ \cline{2-2}
 & TransDeepLab~\cite{azad2022transdeeplab} &  \\ \cline{2-2}
  & MISSForme~\cite{huang2022missformer} &  \\ \cline{1-2}
 
\multirow{5}{*}{Hybrid CNN and Transformer} & TransUNet\cite{chen2021transunet}&  \\ \cline{2-2}
 & HiFormer~\cite{heidari2023hiformer} &  \\ \cline{2-2}
 & R50-ViT~\cite{dosovitskiy2020image} &  \\ \cline{2-2}
  & CS-UNet~\cite{alrfou2024cs} &  \\ \cline{2-2}
 & GPA-TUNet~\cite{li2022gpa} &  \\ \cline{1-2}

\end{tabular}
\label{exper}
\end{table}

\section{Experimental Evaluation}
We compared the performance of GCtx-UNet with state-of-the-art algorithms including CNN-based, Transformer-based, and hybrid algorithms. Table~\ref{exper} summarizes the list of methods that used in our experiments.
We evaluated the performance of GCtx-UNet pre-trained on MedNet and ImageNet on three different types of medical image datasets, including the Synapse multi-organ segmentation dataset (Synapse), Automated Cardiac
Diagnosis Challenge (ACDC)~\cite{bernard2018deep}, and the Polyp datasets which includes CVC-ClinicDB~\cite{bernal2015wm}, Kvasir-SEG~\cite{jha2020kvasir}, CVC-300,
ColonDB~\cite{tajbakhsh2015automated}, and ETIS-LaribDB~\cite{silva2014toward}.

\begin{description}
\item[Synapse] 
Synapse multi-organ segmentation dataset (Synapse)
includes 30 patient cases with 3779 axial abdominal clinical CT images, where 18 cases are used for training and 12 cases are used for testing. The dataset contains 8 abdominal organs (aorta, gallbladder, left kidney, right kidney, liver, pancreas, spleen, and stomach). Each CT volume includes $85 \sim 198$ slices of $512\times512$ pixel images, with a voxel spatial resolution of $[0.54 \sim 0.54] \times [0.98 \sim 0.98] \times [2.5 \sim 5.0]~ mm^3$.

\item[ACDC]
Automated Cardiac Diagnosis Challenge dataset (ACDC)~\cite{bernard2018deep} compiles MRI scan results of various patients from the MICCAI 2017 dataset.
The ACDC dataset contains 100 cardiac MRI scans, each containing three organs: the right ventricle (RV), the myocardium (Myo), and the left ventricle (LV).
Following TransUNet~\cite{chen2021transunet}, we partitioned the dataset into 70 training cases, 10 validation cases, and 20 test cases.

\item[Polyp]
We used 5 polyp datasets with early colorectal cancer diagnosis images. 
The CVC-ClinicDB~\cite{bernal2015wm} and Kvasir-SEG~\cite{jha2020kvasir} datasets are used for binary segmentation. The CVC-ClinicDB dataset contains 612 RGB colonoscopy images with labeled polyps from MICCAI 2015 with a pixel resolution of $288 \times 384$. The Kvasir-SEG dataset contains 1000 polyp images with a pixel resolution ranging from $332 \times 487$ to $1920 \times 1072$ and their corresponding ground truth. Following the setting in PraNet~\cite{fan2020pranet}, we used 900 images from the CVC-ClinicDB dataset and 548 images from the Kvasir dataset for training. The remaining 64 images from CVC-ClinicDB and 100 images from Kvasir were used as test sets. To evaluate the generalization performance, we tested the model on three unseen datasets: CVC-300, CVC-ColonDB, and ETIS-LaribDB.
\end{description}

\subsection{Implementation}
The GCtx-UNet is implemented with PyTorch library and the model is trained on an Nvidia GeForce GTX TITAN X with 12 GB of memory. The input image size was reduced to $224 \times 224$ pixels. Our model is trained with batch size 24, learning rate 0.0001, and AdamW optimizer with momentum 0.9 and weight decay 0.0001. 
For evaluation metrics, we used the average Dice-Similarity Coefficient (DSC) across the Synapse, ADCD, and Polyp datasets. Additionally, the average 95\% Hausdorff Distance (HD) was used on the Synapse dataset.
HD metric provides a more precise estimate of performance with respect to boundary errors. DSC values range from 0 to 1 with the larger values indicating better performance while the smaller values of HD indicate better performance.

\subsection{Synapse}
We compare GCtx-UNet with current state-of-the-art methods on the Synapse dataset, including U-Net, Att-UNet, TransUnet, SwinUnet, MISSFormer, TransDeepLab, HiFormer, GPA-TUNet, and CS-UNet. Table~\ref{table:Med} summarizes the performance comparison in DSC and HD.

GCtx-UNet pre-trained on MedNet has the second-best average DSC (82.39\%) and the third-best average HD (15.94 mm). When pre-trained on ImageNet, GCtx-UNet is the fourth-best in both average DSC (81.95\%) and average HD (16.8 mm)
Note that these metrics are not too far from the best average DSC (83.27\%) and average HD (14.7 mm), which are from CS-UNet and HiFormer, which have much higher computation complexity.
Also note that GCtx-UNet significantly outperforms CNN-based methods UNet and Att-UNet. For instance, UNet has an average DSC of 76.85\% and an average HD of 39.70mm and Att-UNet has an average DSC of 77.77\% and an average HD of 36.02mm. 
GCtx-UNet has bettern performance than Transformer-based methods like Transdeeplab, MISSFormer, and Swin-UNet, which has an average DSC ranging from 79.13 to 81.96\% and an average HD ranging from 18.20 to 21.25 mm. 
GCtx-UNet's performance is also competitive with the hybrid models like TransUnet, GPA-TUNet, HiFormer, and CS-UNet, which have average DSC ranging from 77.48 to 83.27\% and average HD ranging from 14.70 to 31.69 mm.


\begin{table}[ht]
\caption{Comparison of GCtx-UNet and state-of-the-art algorithms on Synapse (the columns are average DSC in \%, average HD in mm, and DSC in \% for each organ). Blue indicates the best result and red displays the second-best.
GCtx-UNet$^1$ is pre-trained on ImageNet while
GCtx-UNet$^2$ is pre-trained on MedNet.
All other models are pre-trained on ImageNet.
}
\begin{tabular}
{|p{2.4cm}|c c| c c c c c c c c|}
\hline
Algorithm & DSC↑ &{ HD↓} & {Aorta} &{Gallbladder}& {Kid(L)} & {Kid(R)} & {Liver} & {Pancreas} & {Spleen} & {Stomach} \\
\hline \hline
U-Net \cite{ronneberger2015u} & 76.85 & 39.70 & \textbf{\textcolor{red}{89.07}} & \textbf{\textcolor{red}{69.72}} & 77.77 & 68.60 & 93.43 & 53.98 & 86.67 & 75.58 \\  
Att-UNet \cite{oktay2018attention} & 77.77 & 36.02 & \textbf{\textcolor{blue}{89.55}} & 68.88 & 77.98 & 71.11 &93.57 & 58.04& 87.30& 75.75\\ \hline 
Swin-UNet \cite{cao2022swin} & 79.13 & 21.55 & 85.47 & 66.53 & 83.2 & 79.61 & 94.29 & 56.58 & 90.66 & 76.60 \\ 
TransDeepLab \cite{azad2022transdeeplab} & 80.16 & 21.25 & 86.04 & 69.16 & 84.08 & 79.88 & 93.53 & 61.19 & 89.00 & 78.40 \\
MISSFormer \cite{huang2022missformer} &81.96 &18.20 & 86.99 & 68.65 & 85.21 & 82.00 & 94.41 &\textbf{\textcolor{blue}{65.67}} &\textbf{\textcolor{blue}{91.92}} & 80.81 \\ \hline
TransUNet \cite{chen2021transunet} & 77.48 & 31.69 & 87.23 & 63.13 & 81.87 & 77.02 & 94.08 & 55.86 & 85.08 & 75.62 \\ 
GPA-TUNet \cite{li2022gpa} & 80.37 &20.55 & 88.74 & 65.63 & 83.51 & 80.37 & \textbf{\textcolor{blue}{94.84}} &63.89 & 87.58 & 78.40 \\ 
HiFormer \cite{heidari2023hiformer} & 80.39 &\textbf{\textcolor{blue} {14.70}} & 86.21 & 65.69 &85.23 & 79.77 & 94.61 & 59.52 & 90.99 & 81.08 \\ 
CS-UNet~\cite{alrfou2024cs} & \textbf{\textcolor{blue}{83.27}} & \textbf{\textcolor{red}{15.26}} & 88.07 & \textbf{\textcolor{blue}{71.32}} & \textbf{\textcolor{blue}{88.00}} & \textbf{\textcolor{blue}{84.38}} & \textbf{\textcolor{red}{94.80}} & \textbf{\textcolor{red}{65.64}} & 89.95 & \textbf{\textcolor{red}{83.81}} \\ \hline
GCtx-UNet$^1$ & 81.95 & 16.80 & 86.96 & 66.26 & \textbf{\textcolor{red}{87.75}} & \textbf{\textcolor{red}{83.86}} & 94.53 & 61.06 & 91.42 & 83.74 \\
GCtx-UNet$^2$ & \textbf{\textcolor{red}{82.39}} & 15.94 & 86.30 & 69.32 & 86.11 & 81.89 & 94.64 & 64.88 & \textbf{\textcolor{red}{91.81}} & \textbf{\textcolor{blue}{84.15}} \\ \hline
\end{tabular}
\label{table:Med}
\end{table}

\begin{figure}[ht!]
\centering
\includegraphics[scale=.48]{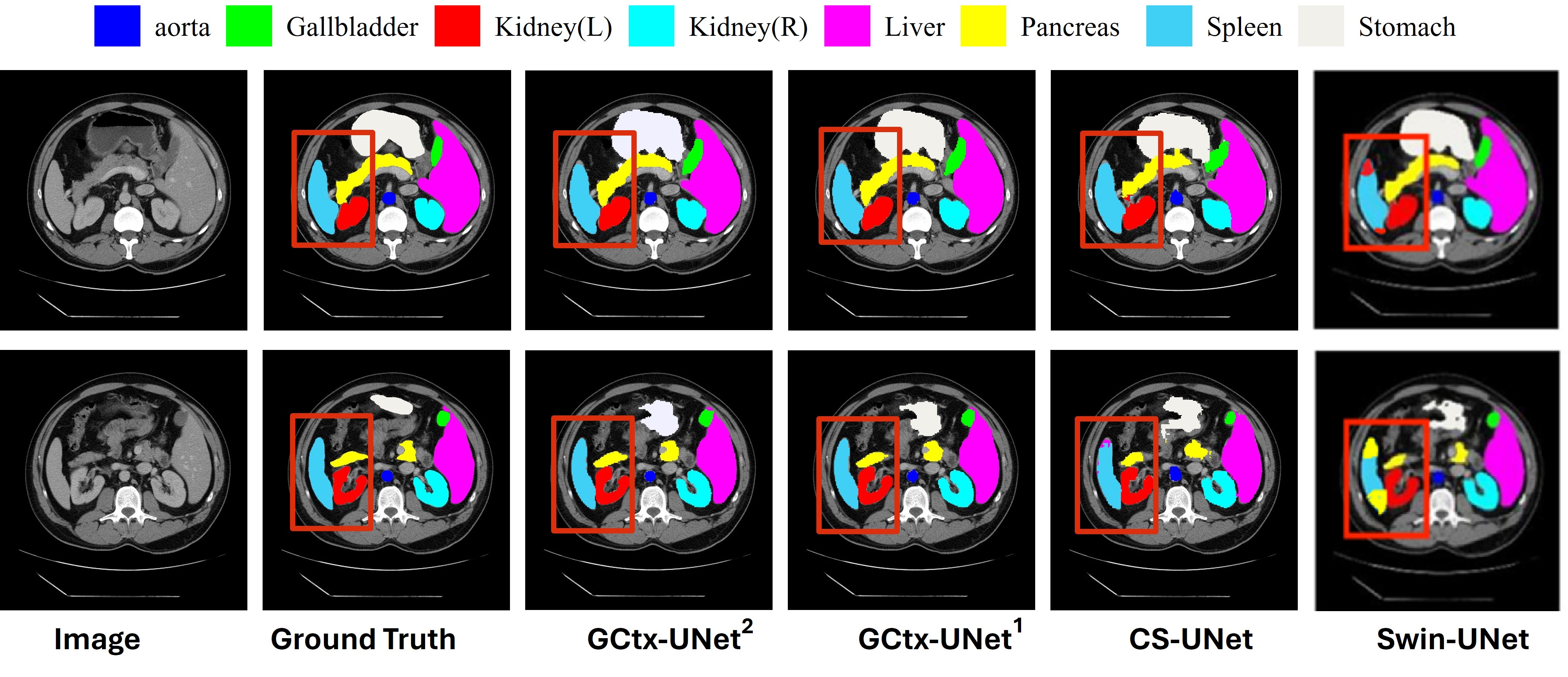}
\caption{Comparison of GCtx-UNet with ground truth, CS-UNet, and Swin-UNet on two sample images in Synapse dataset. Note that GCtx-UNet$^1$ is pre-trained on ImageNet and GCtx-UNet$^2$ is pre-trained on MedNet. The red rectangles identify the regions where Swin-UNet tends to have over-segmentation problems compared to GCtx-UNet and CS-UNet.}
\label{Synap}
\end{figure}

\paragraph{Organ-wise performance}
Figure~\ref{Synap} shows qualitative comparison on two Synapse images, where GCtx-UNet has superior performance than Swin-UNet and CS-UNet. GCtx-UNet segmented most of the organs correctly, with a few misclassifications in the Gallbladder area. In comparison, Swin-UNet over-segmented the spleen (some areas belonging to the spleen were misclassified as the left kidney) and CS-UNet over-segmented the pancreas. We speculate that this improvement is due to the use of GC-ViT, which introduces a parameter-efficient downsampling module with modified Fused MB-Conv blocks. These modifications address the lack of inductive bias in ViTs, enabling GCtx-UNet to accurately capture relatively large regions and perform well with organs close to each other.

The segmentation of larger organs with clear boundaries such as kidney, pancreas, and spleen require the network to capture global features. We speculate that this is why Transformer-based models are more accurate compared to CNN-based models.
The segmentation of smaller organ like aorta benefits more from the detection of local features. This is probably why CNN-based models have more accurate results than Transformed-based models.
The segmentation of larger organs with complex boundaries, such as liver and stomach, requires to capture local and global features. That is probably why hybrid models have more accurate result.
   

\subsection{ACDC}
We compare the performance of GCtx-UNet on the ACDC dataset with some state-of-the-art methods including CNN-based methods (R50-UNet and R50-Atten-UNet), Transformer-based methods (R50-ViT, Swin-UNet, MISSFormer), and hybrid methods (TransUNet, GPA-TUNet, and CS-UNet). The results are shown in Table~\ref{ACDC}.
GCtx-UNet is better than all other methods in terms of average DSC, where GCtx-UNet pre-trained on MedNet has the best performance. 
In particular, GCtx-UNet is better in the Right Ventricle (RV) and Left Ventricle (LV). GCtx-UNet pre-trained on MedNet is also better than GCtx-UNet pre-trained on ImageNet.

\begin{table}[htp]
\caption{Comparison of GCtx-UNet with the state-of-the-art methods on ACDC dataset in DSC. Blue denotes the best results and red denotes the second best.GCtx-UNet$^1$ is pre-trained on ImageNet while GCtx-UNet$^2$ is pre-trained on MedNet.
}\label{ACDC}
\begin{tabular*}{\tblwidth}{@{}LCCCC@{}}
\toprule
Algorithm& DSC(\%)↑& Right Ventricle & Myocardium & Left Ventricle\\ 
\midrule
R50-UNet~\cite{chen2021transunet}&87.55&87.10&80.63&94.92 \\
R50-Atten-UNet~\cite{chen2021transunet}&86.75&87.58&79.20&93.47 \\
\hline
R50-ViT~\cite{chen2021transunet}&87.57&86.07&81.88&94.75 \\
Swin-UNet~\cite{cao2022swin}&90.00&88.55&85.62&95.83\\
MISSFormer~\cite{huang2022missformer}&90.86&89.55&\textbf{\textcolor{blue}{88.04}}&94.99\\
\hline
TransUNet~\cite{chen2021transunet}&89.71&88.86&84.53&95.73\\
GPA-TUNet~\cite{li2022gpa}&90.37&89.44&\textbf{\textcolor{red}{87.98}}&93.68\\
CS-UNet~\cite{alrfou2024cs}&90.38&88.28&86.50&96.35\\
\hline
GCtx-UNet$^1$&\textbf{\textcolor{red}{90.98}}&\textbf{\textcolor{red}{89.63}}&86.77&\textbf{\textcolor{red}{96.55}}\\
GCtx-UNet$^2$&\textbf{\textcolor{blue}{91.23}}&\textbf{\textcolor{blue}{89.88}}&87.25&\textbf{\textcolor{blue}{96.57}}\\
\bottomrule 
\end{tabular*}
\end{table}

\begin{figure}[ht!]
\centering
\includegraphics[scale=.48]{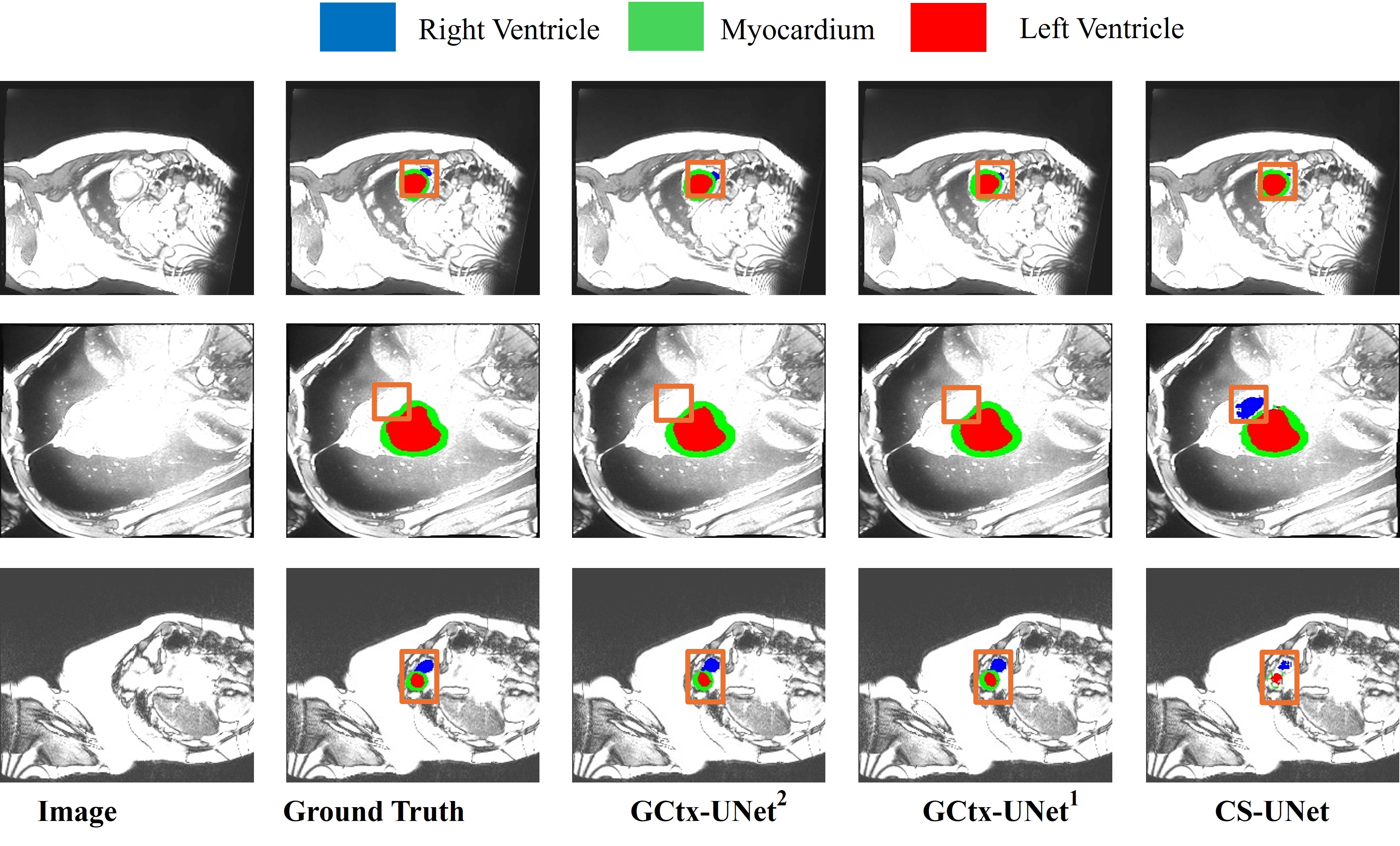}
\caption{Comparison of GCtx-UNet with ground truth and CS-UNet on 3 example images in ACDC dataset. Note that GCtx-UNet$^1$ is pre-trained on ImageNet and GCtx-UNet$^2$ is pre-trained on MedNet. The orange rectangle box identifies the regions where GCtx-UNet$^1$ and CS-UNet have over or under segmentation problems compared to GCtx-UNet$^2$.}
\label{ADCD}
\end{figure}

Figure~\ref{ADCD} includes 3 example images in the ACDC dataset for a qualitative comparison between GCtx-UNet and CS-UNet. GCtx-UNet pre-trained on MetNet was able to segment right ventricle and left ventricle more accurately than CS-UNet and GCtx-UNet pre-trained on ImageNet.

\subsection{Polyp datasets}
We evaluated the performance of GCtx-UNet on several polyp datasets by first training it on 2 seen datasets (CVC-ClinicDB and Kvasir) and then use the trained models on 3 unseen datasets (CVC-ColonDB, ETIS-LaribDB, and CVC-300) to evaluate the generalizability of the models. 
Table~\ref{polyp_data} compares the performance of GCtx-UNet with state-of-the-art CNN-based algorithms (UNet, UNet++, PraNet) and hybrid method (CS-UNet). 
While GCtx-UNet pre-trained on MedNet has the best DSC metric for the Kvasir dataset, it is behind CS-UNet and PraNet on the CVC-ClinicDB dataset though the difference is relatively small.

\paragraph{Generalizability} 
As shown in Table~\ref{polyp_data} (column 3--5), GCtx-UNet is more generalizable to unseen datasets (CVC-ColonDB, ETIS-LaribDB, and CVC-300). Compared to other approaches, GCtx-UNet models have the best and second best DSC value in CVC-ColonDB and ETIS-LaribDB datasets and the second best DSC value in CVC-300 dataset. The overall performance of GCtx-UNet is rather remarkable. 

\begin{table}[htp]
\caption{Comparison of GCtx-UNet and State-Of-The-Art algorithms on Pylop datasets (the columns are average DSC in \%). Blue indicates the best result and red displays the second-best.
GCtx-UNet$^1$ is pre-trained on ImageNet while GCtx-UNet$^2$ is pre-trained on MedNet.
}\label{pylopdata}
\begin{tabular*}{\tblwidth}{@{}L|LL|LLL@{}}
\toprule
\textbf{Methods}&{CVC-ClinicDB}&{Kvasir}&{CVC-300}&{CVC-ColonDB}&{ETIS-LaribDB}\\ 
\midrule
{UNet\cite{ronneberger2015u}} & 82.3 & 81.8 & 71.0 & 51.2  & 39.8\\
{UNet++~\cite{zhou2018unet++}} & 79.4 &82.1  & 70.7  & 48.3  & 40.1  \\
{PraNet~\cite{fan2020pranet}}&\textbf{\textcolor{red}{89.9}}&89.8&\textbf{\textcolor{blue}{87.1}}&70.9&62.8 \\
\hline
{CS-UNet~\cite{alrfou2024cs}}&\textbf{\textcolor{blue}{90.67}}&\textbf{\textcolor{red}{90.00}}&85.59&72.00&64.50\\
\hline
{GCtx-UNet$^1$} &89.48&89.26 &\textbf{\textcolor{red}{86.52}} &\textbf{\textcolor{red}{74.95}}&\textbf{\textcolor{red}{65.97}}\\
{GCtx-UNet$^2$} &89.44&\textbf{\textcolor{blue}{90.02}}&86.20&\textbf{\textcolor{blue}{78.08}}&\textbf{\textcolor{blue}{72.03}} \\
\bottomrule 
\end{tabular*}
\label{polyp_data}
\end{table}


\begin{figure}[ht!]
\centering
\includegraphics[width=\textwidth]{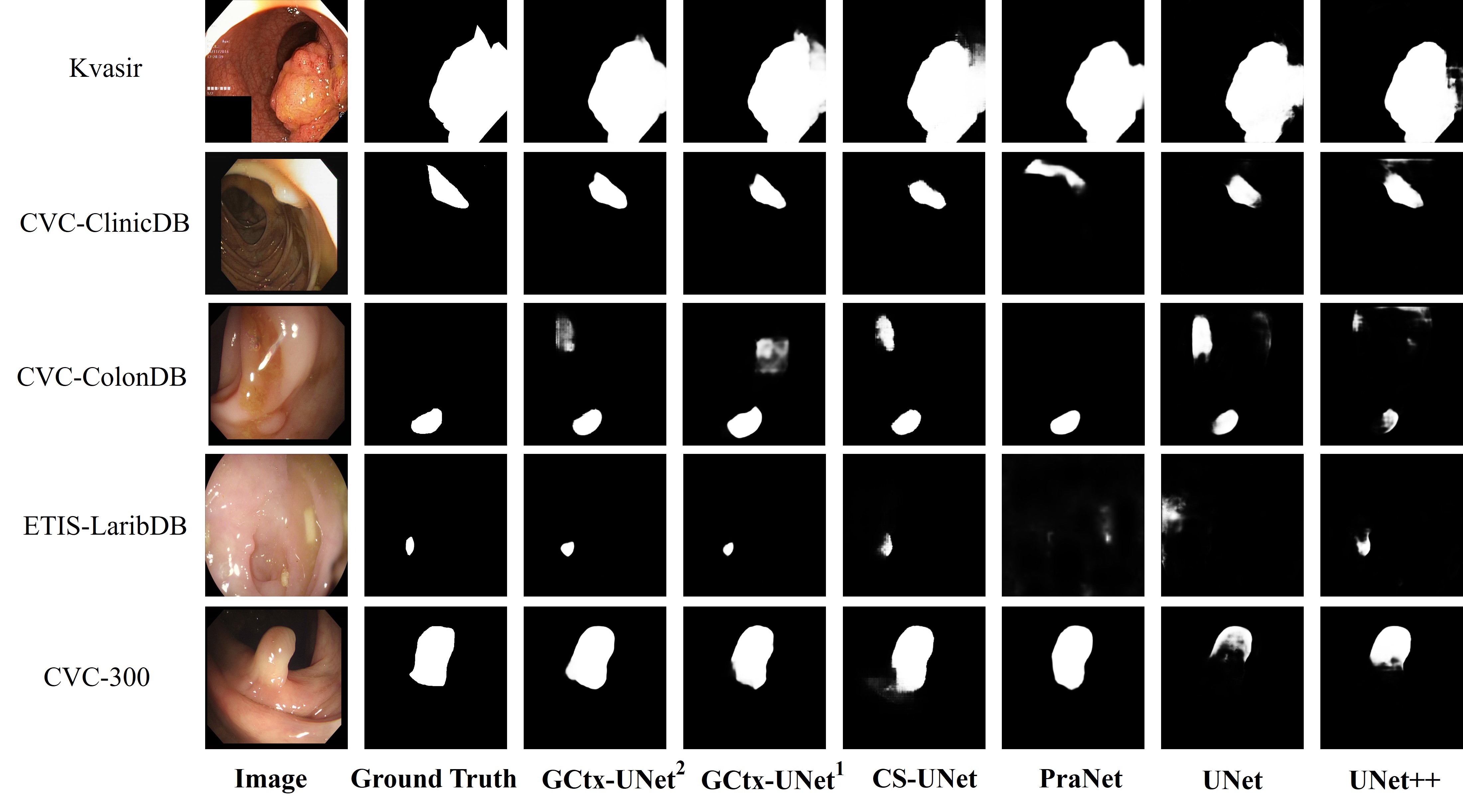}
\caption{Comparison of GCtx-UNet with CS-UNet, PraNet, UNet, and UNet++ on 5 example images in the polyp datasets. GCtx-UNet$^1$ is pre-trained on ImageNet and GCtx-UNet$^2$ is pre-trained on MedNet.}
\label{pylop}
\end{figure}

Figure~\ref{pylop} presents the qualitative segmentation results of various methods, including GCtx-UNet that trained on MedNet and ImageNet. Five samples, one from each dataset, are selected to highlight ambiguous boundaries and small polyps, facilitating a differentiated comparison of segmentation performance. GCtx-UNet, pre-trained on MedNet, shows a significant reduction in false positives and false negatives. This improvement is attributed to its enhanced ability to distinguish between the obscure boundaries of polyp regions and normal regions.

\section{Computation Complexity}
\label{complixtyT}

GCtx-UNet not only has good segmentation performance but also lower computation complexity than the state-of-the-art segmentation methods. We compared the computation complexity of GCtx-UNet with that of Transformer-based and hybrid methods in terms of model parameter numbers, floating point operations (FLOPs) per epoch of training, inference time, and model sizes. This assessment is evaluated on the Synapse dataset. 
As shown in Table~\ref{table:compars}, GCtx-UNet has the least number of parameters at 12.34 million, while the model with nearest number of parameters, TransDeepLab, has 21.14 million parameters. 
Correspondingly, GCtx-UNet has the smallest model size at 49.75 MB while the second smallest model, TransDeepLab, has 86.343 MB.
GCtx-UNet also has the least number of FLOPs per epoch of training at 30.41G while the closest model Swim-UNet has 61.64G. Similarly, GCtx-UNet has the least training time per epoch and highest FPS for inference.

The low computation cost and small memory footprint of GCtx-UNet make it the best trade-off between performance and model complexity. It has better or comparable performance than most of the state-of-the-art models, which have substantially larger model size and higher computation cost. The efficient design of GCtx-UNet highlights its potential for achieving better segmentation results in clinical applications.

\begin{table}[ht!]
\centering
\caption{Comparison of GCtx-UNet, Transformer-based, and hybrid algorithms based on model parameters, the floating point operations (FLOPs) per training epoch, number of epochs to train, training time per epoch (TTPE), and inference time in Frame Per Second (FPS) for 1568 axial abdominal clinical CT images, and model size. 
For FLOPs and TTPE, the batch size is 10. Training epochs is what is needed for the final result. 
}
\begin{tabular}{|c|c|c|c|c|c|c|}
\hline 
Algorithm & \# of params (M) & FLOPs (G) & \# of epochs & TTPE (m:s) & FPS & Model size (MB) \\
\hline 
\hline
TransDeepLab~\cite{azad2022transdeeplab} & 21.14 & 160.00 &200&1:36&20& 86.343 \\
Swin-UNet~\cite{cao2022swin} & 27.17 & 61.64 & 150 & 1:45 & 27 & 108.058 \\ 
MISSFormer~\cite{huang2022missformer} & 42.46 & 98.86 & 400 & 4:01 & 19 & 166.124 \\
\hline
HiFormer~\cite{heidari2023hiformer} & 25.51 & 80.45 & 400 &1:27 & 19 & 101.161 \\
CS-UNet~\cite{alrfou2024cs} & 44.96 & 110.00 & 150 & 2:45 & 26 & 177.613 \\
TransUNet~\cite{chen2021transunet} & 105.28 & 290.00 & 150 & 2:54 & 17 & 414.412 \\
\hline
{GCtx-UNet} & {\bf 12.34} & {\bf 30.41} & 150 & {\bf 1:16} & {\bf 30} & {\bf 49.75} \\
\hline
\end{tabular}
\label{table:compars}
\end{table}


\section{Ablation Study}
To investigate the impact of various factors on model performance, we conducted ablation studies using the Synapse dataset. Below, we discuss the effects of upsampling and the optimal hyperparameters for training our model.

\subsection{Hyper-parameter Tuning}
GCtx-UNet is trained with a combination of two loss functions, dice loss and cross-entropy loss, which aligns with many current segmentation methods.
During the training process, we improve the performance by choosing an optimal combination of dice and cross-entropy losses and an optimal learning rate. We conducted experiments to identify the optimal settings for the combined losses and the learning rate. Table~\ref{hyper} compares the performance of GCtx-UNet in terms of the  Dice-Similarity Coefficient (DSC) and Hausdorff Distance (HD) values for various hyper-parameter values. The optimal DSC and HD are achieved when (dice loss, cross-entropy loss) = (0.3, 0.7) and learning rate is 0.0001. This configuration was used in all of our subsequent experiments.

\begin{table}[htp!]
\caption{Ablation study on the impact of the training hyper-parameters to the performance of GCtx-UNet. The hyper parameters include the loss function (cross entropy loss and dice loss) and learning rate.}
\label{hyper}
\begin{tabular*}{\tblwidth}{@{}LLLLLL@{}}
\toprule
Optimizer & dice loss & cross entropy loss & learning rate & DSC (\%) & HD (mm) \\ 
\midrule
AdamW&0.6&0.4&0.00001&81.67&22.40\\
AdamW&0.6&0.4&0.0001&81.58&22.35\\
AdamW&0.7&0.3&0.00001&81.53&23.16\\
AdamW&0.7&0.3&0.0001&82.39 &15.94\\
\bottomrule
\end{tabular*}
\end{table}

\subsection{Upsampling}
Similar to Swin-UNet~\cite{cao2022swin}, to complement the downsampling layer in the encoder, we specifically designed an upsampling layer in the decoder to perform upsampling and feature dimension increase. To assess the effectiveness of this upsampling layer, we evaluated GCtx-UNet on the Synapse dataset to compare GCtx-Unet with bilinear interpolation (with or without SE (Squeeze and Excitation) block) to GCtx-UNet with transposed convolution (Fused-MBConv module with or without SE block) in the upsampling layer. The results in Table~\ref{upsamp} indicate that GCtx-UNet with transposed convolution (Fused-MBConv module with SE block) in the upsampling layer has the best performance.

\begin{table}[ht!]
\caption{Ablation study on the impact of the upsampling types to the performance of GCtx-UNet.}\label{upsamp}
\begin{tabular*}{\tblwidth}{@{}LLL@{}}
\toprule
Upsampling Type & DSC (\%) & HD (mm) \\ 
\midrule
bilinear interpolation & 80.31 & 22.64\\
bilinear interpolation + SE & 80.89 & 25.76 \\
transposed convolution (Fused-MBConv) & 81.80 & 21.12\\
transposed convolution (Fused-MBConv + SE) & 82.39 & 15.94 \\
\bottomrule
\end{tabular*}
\end{table}

\section{Conclusion}
We introduced GCtx-UNet, a U-shaped network that incorporates a lightweight vision transformer to enhance medical image segmentation by effectively capturing both global and local features. The downsampling and upsampling blocks between encoder and decoder components provide inductive bias and model inter-channel dependencies effectively. 
GCtx-UNet has better or comparable performance than traditional CNN-based, Transformer-based, and hybrid methods on various medical image datasets. At the same time, GCtx-UNet has lower model complexity with less number of model parameters, lower model size, lower training and inference time, and lower FLOPs for training.
The ability of GCtx-UNet to model long-range spatial dependencies and its competitive performance in segmenting complex and small anatomical structures make it a promising tool for clinical applications. 
The architecture's design, which includes GC-ViT encoders and decoders with skip connections, contributes to its high performance while maintaining a lower computational complexity compared to state-of-the-art methods. The pre-training on a medical image dataset -- MedNet and the subsequent evaluation on multiple medical imaging tasks shows the model's robustness and generalization capabilities, positioning GCtx-UNet as a practical and powerful approach for medical image segmentation.
As future work, we plan to introduce GCtx-UNet 3D model for voxel segmentation of medical images.

\section*{Funding statement}
This research was partially funded by the Northwestern Mutual Data Science Institute.


\printcredits

\section*{Declaration of competing interest}
The authors declare that they have no known competing financial interests or personal relationships that could have appeared to influence the work reported in this paper.
\bibliographystyle{cas-model2-names}

\bibliography{cas}

\begin{thebibliography}{35}
\expandafter\ifx\csname natexlab\endcsname\relax\def\natexlab#1{#1}\fi
\providecommand{\url}[1]{\texttt{#1}}
\providecommand{\href}[2]{#2}
\providecommand{\path}[1]{#1}
\providecommand{\DOIprefix}{doi:}
\providecommand{\ArXivprefix}{arXiv:}
\providecommand{\URLprefix}{URL: }
\providecommand{\Pubmedprefix}{pmid:}
\providecommand{\doi}[1]{\href{http://dx.doi.org/#1}{\path{#1}}}
\providecommand{\Pubmed}[1]{\href{pmid:#1}{\path{#1}}}
\providecommand{\bibinfo}[2]{#2}
\ifx\xfnm\relax \def\xfnm[#1]{\unskip,\space#1}\fi
\bibitem[{Alrfou et~al.(2024)Alrfou, Zhao and Kordijazi}]{alrfou2024cs}
\bibinfo{author}{Alrfou, K.}, \bibinfo{author}{Zhao, T.}, \bibinfo{author}{Kordijazi, A.}, \bibinfo{year}{2024}.
\newblock \bibinfo{title}{{CS-UNet}: A generalizable and flexible segmentation algorithm}.
\newblock \bibinfo{journal}{Multimedia Tools and Applications} , \bibinfo{pages}{1--28}.
\bibitem[{Azad et~al.(2022)Azad, Heidari, Shariatnia, Aghdam, Karimijafarbigloo, Adeli and Merhof}]{azad2022transdeeplab}
\bibinfo{author}{Azad, R.}, \bibinfo{author}{Heidari, M.}, \bibinfo{author}{Shariatnia, M.}, \bibinfo{author}{Aghdam, E.K.}, \bibinfo{author}{Karimijafarbigloo, S.}, \bibinfo{author}{Adeli, E.}, \bibinfo{author}{Merhof, D.}, \bibinfo{year}{2022}.
\newblock \bibinfo{title}{Transdeeplab: Convolution-free transformer-based deeplab v3+ for medical image segmentation}, in: \bibinfo{booktitle}{Predictive Intelligence in Medicine: 5th International Workshop, PRIME 2022, Held in Conjunction with MICCAI 2022, Singapore, September 22, 2022, Proceedings}, \bibinfo{organization}{Springer}. pp. \bibinfo{pages}{91--102}.
\newblock \DOIprefix\doi{10.1007/978-3-031-16919-9_9}.
\bibitem[{Azad et~al.(2023)Azad, Jia, Aghdam, Cohen-Adad and Merhof}]{azad2023enhancing}
\bibinfo{author}{Azad, R.}, \bibinfo{author}{Jia, Y.}, \bibinfo{author}{Aghdam, E.K.}, \bibinfo{author}{Cohen-Adad, J.}, \bibinfo{author}{Merhof, D.}, \bibinfo{year}{2023}.
\newblock \bibinfo{title}{Enhancing medical image segmentation with transception: a multi-scale feature fusion approach}.
\newblock \bibinfo{journal}{arXiv preprint arXiv:2301.10847} .
\bibitem[{Bernal et~al.(2015)Bernal, S{\'a}nchez, Fern{\'a}ndez-Esparrach, Gil, Rodr{\'\i}guez and Vilari{\~n}o}]{bernal2015wm}
\bibinfo{author}{Bernal, J.}, \bibinfo{author}{S{\'a}nchez, F.J.}, \bibinfo{author}{Fern{\'a}ndez-Esparrach, G.}, \bibinfo{author}{Gil, D.}, \bibinfo{author}{Rodr{\'\i}guez, C.}, \bibinfo{author}{Vilari{\~n}o, F.}, \bibinfo{year}{2015}.
\newblock \bibinfo{title}{Wm-dova maps for accurate polyp highlighting in colonoscopy: Validation vs. saliency maps from physicians}.
\newblock \bibinfo{journal}{Computerized medical imaging and graphics} \bibinfo{volume}{43}, \bibinfo{pages}{99--111}.
\bibitem[{Bernard et~al.(2018)Bernard, Lalande, Zotti, Cervenansky, Yang, Heng, Cetin, Lekadir, Camara, Ballester et~al.}]{bernard2018deep}
\bibinfo{author}{Bernard, O.}, \bibinfo{author}{Lalande, A.}, \bibinfo{author}{Zotti, C.}, \bibinfo{author}{Cervenansky, F.}, \bibinfo{author}{Yang, X.}, \bibinfo{author}{Heng, P.A.}, \bibinfo{author}{Cetin, I.}, \bibinfo{author}{Lekadir, K.}, \bibinfo{author}{Camara, O.}, \bibinfo{author}{Ballester, M.A.G.}, et~al., \bibinfo{year}{2018}.
\newblock \bibinfo{title}{Deep learning techniques for automatic mri cardiac multi-structures segmentation and diagnosis: is the problem solved?}
\newblock \bibinfo{journal}{IEEE transactions on medical imaging} \bibinfo{volume}{37}, \bibinfo{pages}{2514--2525}.
\bibitem[{Cao et~al.(2022)Cao, Wang, Chen, Jiang, Zhang, Tian and Wang}]{cao2022swin}
\bibinfo{author}{Cao, H.}, \bibinfo{author}{Wang, Y.}, \bibinfo{author}{Chen, J.}, \bibinfo{author}{Jiang, D.}, \bibinfo{author}{Zhang, X.}, \bibinfo{author}{Tian, Q.}, \bibinfo{author}{Wang, M.}, \bibinfo{year}{2022}.
\newblock \bibinfo{title}{Swin-unet: Unet-like pure transformer for medical image segmentation}, in: \bibinfo{booktitle}{European Conference on Computer Vision}, \bibinfo{organization}{Springer}. pp. \bibinfo{pages}{205--218}.
\newblock \DOIprefix\doi{10.1007/978-3-031-25066-8_9}.
\bibitem[{Chang et~al.(2021)Chang, Menghan, Guangtao and Xiao-Ping}]{chang2021transclaw}
\bibinfo{author}{Chang, Y.}, \bibinfo{author}{Menghan, H.}, \bibinfo{author}{Guangtao, Z.}, \bibinfo{author}{Xiao-Ping, Z.}, \bibinfo{year}{2021}.
\newblock \bibinfo{title}{Transclaw u-net: Claw u-net with transformers for medical image segmentation}.
\newblock \bibinfo{journal}{arXiv preprint arXiv:2107.05188} .
\bibitem[{Chen et~al.(2021)Chen, Lu, Yu, Luo, Adeli, Wang, Lu, Yuille and Zhou}]{chen2021transunet}
\bibinfo{author}{Chen, J.}, \bibinfo{author}{Lu, Y.}, \bibinfo{author}{Yu, Q.}, \bibinfo{author}{Luo, X.}, \bibinfo{author}{Adeli, E.}, \bibinfo{author}{Wang, Y.}, \bibinfo{author}{Lu, L.}, \bibinfo{author}{Yuille, A.L.}, \bibinfo{author}{Zhou, Y.}, \bibinfo{year}{2021}.
\newblock \bibinfo{title}{Transunet: Transformers make strong encoders for medical image segmentation}.
\newblock \bibinfo{journal}{arXiv preprint arXiv:2102.04306} \DOIprefix\doi{10.48550/arXiv.2102.04306}.
\bibitem[{Dosovitskiy et~al.(2020)Dosovitskiy, Beyer, Kolesnikov, Weissenborn, Zhai, Unterthiner, Dehghani, Minderer, Heigold, Gelly et~al.}]{dosovitskiy2020image}
\bibinfo{author}{Dosovitskiy, A.}, \bibinfo{author}{Beyer, L.}, \bibinfo{author}{Kolesnikov, A.}, \bibinfo{author}{Weissenborn, D.}, \bibinfo{author}{Zhai, X.}, \bibinfo{author}{Unterthiner, T.}, \bibinfo{author}{Dehghani, M.}, \bibinfo{author}{Minderer, M.}, \bibinfo{author}{Heigold, G.}, \bibinfo{author}{Gelly, S.}, et~al., \bibinfo{year}{2020}.
\newblock \bibinfo{title}{An image is worth 16x16 words: Transformers for image recognition at scale}.
\newblock \bibinfo{journal}{arXiv preprint arXiv:2010.11929} \DOIprefix\doi{10.48550/arXiv.2010.11929}.
\bibitem[{Fan et~al.(2020)Fan, Ji, Zhou, Chen, Fu, Shen and Shao}]{fan2020pranet}
\bibinfo{author}{Fan, D.P.}, \bibinfo{author}{Ji, G.P.}, \bibinfo{author}{Zhou, T.}, \bibinfo{author}{Chen, G.}, \bibinfo{author}{Fu, H.}, \bibinfo{author}{Shen, J.}, \bibinfo{author}{Shao, L.}, \bibinfo{year}{2020}.
\newblock \bibinfo{title}{Pranet: Parallel reverse attention network for polyp segmentation}, in: \bibinfo{booktitle}{International conference on medical image computing and computer-assisted intervention}, \bibinfo{organization}{Springer}. pp. \bibinfo{pages}{263--273}.
\bibitem[{Fu et~al.(2022)Fu, Li and Hua}]{fu2022deau}
\bibinfo{author}{Fu, Z.}, \bibinfo{author}{Li, J.}, \bibinfo{author}{Hua, Z.}, \bibinfo{year}{2022}.
\newblock \bibinfo{title}{Deau-net: Attention networks based on dual encoder for medical image segmentation}.
\newblock \bibinfo{journal}{Computers in Biology and Medicine} \bibinfo{volume}{150}, \bibinfo{pages}{106197}.
\bibitem[{Hatamizadeh et~al.(2022)Hatamizadeh, Tang, Nath, Yang, Myronenko, Landman, Roth and Xu}]{hatamizadeh2022unetr}
\bibinfo{author}{Hatamizadeh, A.}, \bibinfo{author}{Tang, Y.}, \bibinfo{author}{Nath, V.}, \bibinfo{author}{Yang, D.}, \bibinfo{author}{Myronenko, A.}, \bibinfo{author}{Landman, B.}, \bibinfo{author}{Roth, H.R.}, \bibinfo{author}{Xu, D.}, \bibinfo{year}{2022}.
\newblock \bibinfo{title}{Unetr: Transformers for 3d medical image segmentation}, in: \bibinfo{booktitle}{Proceedings of the IEEE/CVF winter conference on applications of computer vision}, pp. \bibinfo{pages}{574--584}.
\bibitem[{Hatamizadeh et~al.(2023)Hatamizadeh, Yin, Heinrich, Kautz and Molchanov}]{hatamizadeh2023global}
\bibinfo{author}{Hatamizadeh, A.}, \bibinfo{author}{Yin, H.}, \bibinfo{author}{Heinrich, G.}, \bibinfo{author}{Kautz, J.}, \bibinfo{author}{Molchanov, P.}, \bibinfo{year}{2023}.
\newblock \bibinfo{title}{Global context vision transformers}, in: \bibinfo{booktitle}{International Conference on Machine Learning}, \bibinfo{organization}{PMLR}. pp. \bibinfo{pages}{12633--12646}.
\bibitem[{Heidari et~al.(2023)Heidari, Kazerouni, Soltany, Azad, Aghdam, Cohen-Adad and Merhof}]{heidari2023hiformer}
\bibinfo{author}{Heidari, M.}, \bibinfo{author}{Kazerouni, A.}, \bibinfo{author}{Soltany, M.}, \bibinfo{author}{Azad, R.}, \bibinfo{author}{Aghdam, E.K.}, \bibinfo{author}{Cohen-Adad, J.}, \bibinfo{author}{Merhof, D.}, \bibinfo{year}{2023}.
\newblock \bibinfo{title}{Hiformer: Hierarchical multi-scale representations using transformers for medical image segmentation}, in: \bibinfo{booktitle}{Proceedings of the IEEE/CVF Winter Conference on Applications of Computer Vision}, pp. \bibinfo{pages}{6202--6212}.
\newblock \DOIprefix\doi{WACV56688.2023.00614}.
\bibitem[{Huang et~al.(2020)Huang, Lin, Tong, Hu, Zhang, Iwamoto, Han, Chen and Wu}]{huang2020unet}
\bibinfo{author}{Huang, H.}, \bibinfo{author}{Lin, L.}, \bibinfo{author}{Tong, R.}, \bibinfo{author}{Hu, H.}, \bibinfo{author}{Zhang, Q.}, \bibinfo{author}{Iwamoto, Y.}, \bibinfo{author}{Han, X.}, \bibinfo{author}{Chen, Y.W.}, \bibinfo{author}{Wu, J.}, \bibinfo{year}{2020}.
\newblock \bibinfo{title}{Unet 3+: A full-scale connected unet for medical image segmentation}, in: \bibinfo{booktitle}{ICASSP 2020-2020 IEEE international conference on acoustics, speech and signal processing (ICASSP)}, \bibinfo{organization}{IEEE}. pp. \bibinfo{pages}{1055--1059}.
\bibitem[{Huang et~al.(2022)Huang, Deng, Li, Yuan and Fu}]{huang2022missformer}
\bibinfo{author}{Huang, X.}, \bibinfo{author}{Deng, Z.}, \bibinfo{author}{Li, D.}, \bibinfo{author}{Yuan, X.}, \bibinfo{author}{Fu, Y.}, \bibinfo{year}{2022}.
\newblock \bibinfo{title}{Missformer: an effective transformer for 2d medical image segmentation}.
\newblock \bibinfo{journal}{IEEE transactions on medical imaging} .
\bibitem[{Jha et~al.(2020)Jha, Smedsrud, Riegler, Halvorsen, De~Lange, Johansen and Johansen}]{jha2020kvasir}
\bibinfo{author}{Jha, D.}, \bibinfo{author}{Smedsrud, P.H.}, \bibinfo{author}{Riegler, M.A.}, \bibinfo{author}{Halvorsen, P.}, \bibinfo{author}{De~Lange, T.}, \bibinfo{author}{Johansen, D.}, \bibinfo{author}{Johansen, H.D.}, \bibinfo{year}{2020}.
\newblock \bibinfo{title}{Kvasir-seg: A segmented polyp dataset}, in: \bibinfo{booktitle}{MultiMedia Modeling: 26th International Conference, MMM 2020, Daejeon, South Korea, January 5--8, 2020, Proceedings, Part II 26}, \bibinfo{organization}{Springer}. pp. \bibinfo{pages}{451--462}.
\bibitem[{Kingma and Ba(2014)}]{kingma2014adam}
\bibinfo{author}{Kingma, D.P.}, \bibinfo{author}{Ba, J.}, \bibinfo{year}{2014}.
\newblock \bibinfo{title}{Adam: A method for stochastic optimization}.
\newblock \bibinfo{journal}{arXiv preprint arXiv:1412.6980} .
\bibitem[{Li et~al.(2024)Li, Wang and Cheng}]{li2024enhanced}
\bibinfo{author}{Li, C.}, \bibinfo{author}{Wang, L.}, \bibinfo{author}{Cheng, S.}, \bibinfo{year}{2024}.
\newblock \bibinfo{title}{Enhanced transformer encoder and hybrid cascaded upsampler for medical image segmentation}.
\newblock \bibinfo{journal}{Expert Systems with Applications} \bibinfo{volume}{238}, \bibinfo{pages}{121965}.
\bibitem[{Li et~al.(2022)Li, Wang and Li}]{li2022gpa}
\bibinfo{author}{Li, C.}, \bibinfo{author}{Wang, L.}, \bibinfo{author}{Li, Y.}, \bibinfo{year}{2022}.
\newblock \bibinfo{title}{Gpa-tunet: Transformer and gpa attention co-encoder for medical image segmentation} .
\bibitem[{Liu et~al.(2021)Liu, Lin, Cao, Hu, Wei, Zhang, Lin and Guo}]{liu2021swin}
\bibinfo{author}{Liu, Z.}, \bibinfo{author}{Lin, Y.}, \bibinfo{author}{Cao, Y.}, \bibinfo{author}{Hu, H.}, \bibinfo{author}{Wei, Y.}, \bibinfo{author}{Zhang, Z.}, \bibinfo{author}{Lin, S.}, \bibinfo{author}{Guo, B.}, \bibinfo{year}{2021}.
\newblock \bibinfo{title}{Swin transformer: Hierarchical vision transformer using shifted windows}, in: \bibinfo{booktitle}{Proceedings of the IEEE/CVF international conference on computer vision}, pp. \bibinfo{pages}{10012--10022}.
\newblock \DOIprefix\doi{10.1109/ICCV48922.2021.00986}.
\bibitem[{No Author(b)}]{bib12}
No Author, \bibinfo{year}{2020}b.
\newblock \bibinfo{title}{"kaggle:ct scan"}.
\newblock \bibinfo{note}{\url{https://www.kaggle.com/datasets/mohamedhanyyy/chest-ctscan-images}}.
\bibitem[{No Author(a)}]{bib13}
No Author, \bibinfo{year}{2021}a.
\newblock \bibinfo{title}{"kaggle:ct kidney dataset"}.
\newblock \bibinfo{note}{\url{https://www.kaggle.com/datasets/nazmul0087/ct-kidney-dataset-normal-cyst-tumor-and-stone/data}}.
\bibitem[{No Author(c)}]{bib14}
No Author, \bibinfo{year}{2024}c.
\newblock \bibinfo{title}{"kaggle:medical scan classification dataset"}.
\newblock \bibinfo{note}{\url{https://www.kaggle.com/datasets/arjunbasandrai/medical-scan-classification-dataset}}.
\bibitem[{Oktay et~al.(2018)Oktay, Schlemper, Folgoc, Lee, Heinrich, Misawa, Mori, McDonagh, Hammerla, Kainz et~al.}]{oktay2018attention}
\bibinfo{author}{Oktay, O.}, \bibinfo{author}{Schlemper, J.}, \bibinfo{author}{Folgoc, L.L.}, \bibinfo{author}{Lee, M.}, \bibinfo{author}{Heinrich, M.}, \bibinfo{author}{Misawa, K.}, \bibinfo{author}{Mori, K.}, \bibinfo{author}{McDonagh, S.}, \bibinfo{author}{Hammerla, N.Y.}, \bibinfo{author}{Kainz, B.}, et~al., \bibinfo{year}{2018}.
\newblock \bibinfo{title}{Attention u-net: Learning where to look for the pancreas}.
\newblock \bibinfo{journal}{arXiv preprint arXiv:1804.03999} \DOIprefix\doi{10.48550/arXiv.1804.03999}.
\bibitem[{Ronneberger et~al.(2015)Ronneberger, Fischer and Brox}]{ronneberger2015u}
\bibinfo{author}{Ronneberger, O.}, \bibinfo{author}{Fischer, P.}, \bibinfo{author}{Brox, T.}, \bibinfo{year}{2015}.
\newblock \bibinfo{title}{U-net: Convolutional networks for biomedical image segmentation}, in: \bibinfo{booktitle}{Medical Image Computing and Computer-Assisted Intervention--MICCAI 2015: 18th International Conference, Munich, Germany, October 5-9, 2015, Proceedings, Part III 18}, \bibinfo{organization}{Springer}. pp. \bibinfo{pages}{234--241}.
\newblock \DOIprefix\doi{10.1007/978-3-319-24574-4_28}.
\bibitem[{Shamshad et~al.(2023)Shamshad, Khan, Zamir, Khan, Hayat, Khan and Fu}]{shamshad2023transformers}
\bibinfo{author}{Shamshad, F.}, \bibinfo{author}{Khan, S.}, \bibinfo{author}{Zamir, S.W.}, \bibinfo{author}{Khan, M.H.}, \bibinfo{author}{Hayat, M.}, \bibinfo{author}{Khan, F.S.}, \bibinfo{author}{Fu, H.}, \bibinfo{year}{2023}.
\newblock \bibinfo{title}{Transformers in medical imaging: A survey}.
\newblock \bibinfo{journal}{Medical Image Analysis} , \bibinfo{pages}{102802}.
\bibitem[{Silva et~al.(2014)Silva, Histace, Romain, Dray and Granado}]{silva2014toward}
\bibinfo{author}{Silva, J.}, \bibinfo{author}{Histace, A.}, \bibinfo{author}{Romain, O.}, \bibinfo{author}{Dray, X.}, \bibinfo{author}{Granado, B.}, \bibinfo{year}{2014}.
\newblock \bibinfo{title}{Toward embedded detection of polyps in wce images for early diagnosis of colorectal cancer}.
\newblock \bibinfo{journal}{International journal of computer assisted radiology and surgery} \bibinfo{volume}{9}, \bibinfo{pages}{283--293}.
\bibitem[{Stuckner et~al.(2022)Stuckner, Harder and Smith}]{stuckner2022microstructure}
\bibinfo{author}{Stuckner, J.}, \bibinfo{author}{Harder, B.}, \bibinfo{author}{Smith, T.M.}, \bibinfo{year}{2022}.
\newblock \bibinfo{title}{Microstructure segmentation with deep learning encoders pre-trained on a large microscopy dataset}.
\newblock \bibinfo{journal}{npj Computational Materials} \bibinfo{volume}{8}, \bibinfo{pages}{200}.
\bibitem[{Subramoniam et~al.(2022)Subramoniam, Aparna, Anurenjan and Sreeni}]{subramoniam2022deep}
\bibinfo{author}{Subramoniam, M.}, \bibinfo{author}{Aparna, T.}, \bibinfo{author}{Anurenjan, P.}, \bibinfo{author}{Sreeni, K.}, \bibinfo{year}{2022}.
\newblock \bibinfo{title}{Deep learning-based prediction of alzheimer’s disease from magnetic resonance images}, in: \bibinfo{booktitle}{Intelligent vision in healthcare}. \bibinfo{publisher}{Springer}, pp. \bibinfo{pages}{145--151}.
\bibitem[{Tajbakhsh et~al.(2015)Tajbakhsh, Gurudu and Liang}]{tajbakhsh2015automated}
\bibinfo{author}{Tajbakhsh, N.}, \bibinfo{author}{Gurudu, S.R.}, \bibinfo{author}{Liang, J.}, \bibinfo{year}{2015}.
\newblock \bibinfo{title}{Automated polyp detection in colonoscopy videos using shape and context information}.
\newblock \bibinfo{journal}{IEEE transactions on medical imaging} \bibinfo{volume}{35}, \bibinfo{pages}{630--644}.
\bibitem[{Xiao et~al.(2018)Xiao, Lian, Luo and Li}]{xiao2018weighted}
\bibinfo{author}{Xiao, X.}, \bibinfo{author}{Lian, S.}, \bibinfo{author}{Luo, Z.}, \bibinfo{author}{Li, S.}, \bibinfo{year}{2018}.
\newblock \bibinfo{title}{Weighted res-unet for high-quality retina vessel segmentation}, in: \bibinfo{booktitle}{2018 9th international conference on information technology in medicine and education (ITME)}, \bibinfo{organization}{IEEE}. pp. \bibinfo{pages}{327--331}.
\bibitem[{Zhang et~al.(2023)Zhang, Lu, Zhao, Hu, Su and Yuan}]{zhang2023accpg}
\bibinfo{author}{Zhang, W.}, \bibinfo{author}{Lu, F.}, \bibinfo{author}{Zhao, W.}, \bibinfo{author}{Hu, Y.}, \bibinfo{author}{Su, H.}, \bibinfo{author}{Yuan, M.}, \bibinfo{year}{2023}.
\newblock \bibinfo{title}{Accpg-net: A skin lesion segmentation network with adaptive channel-context-aware pyramid attention and global feature fusion}.
\newblock \bibinfo{journal}{Computers in Biology and Medicine} \bibinfo{volume}{154}, \bibinfo{pages}{106580}.
\bibitem[{Zhou et~al.(2018)Zhou, Rahman~Siddiquee, Tajbakhsh and Liang}]{zhou2018unet++}
\bibinfo{author}{Zhou, Z.}, \bibinfo{author}{Rahman~Siddiquee, M.M.}, \bibinfo{author}{Tajbakhsh, N.}, \bibinfo{author}{Liang, J.}, \bibinfo{year}{2018}.
\newblock \bibinfo{title}{Unet++: A nested u-net architecture for medical image segmentation}, in: \bibinfo{booktitle}{Deep Learning in Medical Image Analysis and Multimodal Learning for Clinical Decision Support: 4th International Workshop, DLMIA 2018, and 8th International Workshop, ML-CDS 2018, Held in Conjunction with MICCAI 2018, Granada, Spain, September 20, 2018, Proceedings 4}, \bibinfo{organization}{Springer}. pp. \bibinfo{pages}{3--11}.
\bibitem[{Zhou et~al.(2019)Zhou, Siddiquee, Tajbakhsh and Liang}]{zhou2019unet++}
\bibinfo{author}{Zhou, Z.}, \bibinfo{author}{Siddiquee, M.M.R.}, \bibinfo{author}{Tajbakhsh, N.}, \bibinfo{author}{Liang, J.}, \bibinfo{year}{2019}.
\newblock \bibinfo{title}{Unet++: Redesigning skip connections to exploit multiscale features in image segmentation}.
\newblock \bibinfo{journal}{IEEE transactions on medical imaging} \bibinfo{volume}{39}, \bibinfo{pages}{1856--1867}.

\end{thebibliography}

\bio{}
\endbio


\end{document}